\documentstyle[12pt]{article}
\newcommand{\ep}{\varepsilon}
\newcommand{\vf}{\varphi}
\newcommand{\al}{\alpha}
\newcommand{\be}{\beta}
\newcommand{\ga}{\gamma}
\newcommand{\si}{\sigma}
\newcommand{\de}{\delta}

\newcommand{\bX}{{\bf X}}
\newcommand{\bH}{{\bf H}}
\newcommand{\bV}{{\bf V}}
\newcommand{\bR}{{\bf R}}
\begin{document}
\title{Spectral Properties of Faddeev Equations in Differential Form} 
\author{S.L. Yakovlev\thanks{{\it E-mail address:}  
 yakovlev@mph.phys.spbu.ru
}}
\date{ }
\maketitle 
\begin{center}
{\it Department of Mathematical and Computational
Physics, St. Petersburg State University,  
 198904 St. Petersburg, Petrodvoretz, Ulyanovskaya Str.
1, Russia}  
\end{center}
\vskip 1cm 
\section{Introduction}
Faddeev equations in differential form were introduced by H.P. Noyes and 
H.~Fiedelday in 1968 \cite{NF}
\begin{equation}
(H_0-E)\vf_{\al}+V_{\al}\sum_{\be=1}^{3}\vf_{\be}=0,
\label{fadeq}
\end{equation}
and since that time are used extensively as for investigating 
theoretical aspects of the three-body problem as well as for numerical
solutions of three-body bound-state and scattering state problems.
The simple formula 
$$
\sum_{\be=1}^{3}\vf_{\be}=\Psi
$$
allows one to obtain the solution to the three-body Schr\"odinger equation
$$
(H_0+\sum_{\be=1}^{3}V_{\be}-E)\Psi=0
$$
in the case when 
\begin{equation}
\sum_{\be=1}^{3}\vf_{\be} \ne 0.
\label{nonzero}
\end{equation}
Such solutions of (\ref{fadeq}) can be called {\bf physical}. The proper
asymptotic boundary conditions should be added to Eqs. (\ref{fadeq}) in
order to guarantee (\ref{nonzero}). This conditions were studied by many
authors and are well known \cite{FM}. So that, I will not discuss them
here. 

On the other hand, Eqs. (\ref{fadeq}) themselves allow solutions of the
different type (to physical ones) with the property
$$
\sum_{\be=1}^{3}\vf_{\be}=0.
$$
This solutions can be constructed explicitly and have the form
$$
\vf_{\al}=\si_{\al}\phi^{0},
$$
where $\phi^{0}$ is an eigenfunction of operator $H_0$:
$$
H_{0}\phi^{0}=E^{0}\phi^{0}
$$
and $\si_{\al}$, $\al=1,2,3$ are numbers such that
$\sum_{\al=1}^{3}\si_{\al}=0$. 
The solutions of this type can be called {\bf spurious} or {\bf ghost},
because they do not correspond to any three-body system and do not contain
any information about interactions between particles. First observation of
the existence of spurious solutions was made in ref. \cite{Friar}. Some
spurious solutions corresponding to particular values of the total angular
momentum were found in refs. \cite{Pup1}, \cite{Pup2}. All the spurious
solutions on 
subspaces with fixed total angular momentum were constructed in ref.
\cite{RYa}.  

So that, there exist at least two types of solutions to Eqs. 
(\ref{fadeq}) corresponding to real energy:\\ 
\hspace*{1cm} {\bf physical} ones with the property
$\sum\limits_{\be=1}^{3}\vf_{\be}\ne 0$, \\
\hspace*{1cm} {\bf spurious} ones with the property
$\sum\limits_{\be=1}^{3}\vf_{\be}= 0$. \\
The QUESTION is do these solutions form the complete set or there could be
exist solutions of different type which do not belong to physical and
spurious classes.
The ANSWER is not so evident because the operator corresponding to Eqs.
(\ref{fadeq}) is not selfadjoint and moreover even symmetrical:
\begin{equation}
\bH= 
\left( 
\begin{array}{ccc}
H_0 & 0    & 0  \\
0   &H_{0} & 0  \\
0   &0     &H_0 \\
\end{array}
\right) + 
\left( 
\begin{array}{ccc}
V_1  & 0    & 0  \\
0    &V_2    & 0  \\
0   &0     &V_3   \\
\end{array}
\right)
\left( 
\begin{array}{ccc}
1 & 1    & 1  \\
1 & 1    & 1  \\
1 & 1    & 1 \\
\end{array}
\right)= \bH_{0}+\bV\bX,
\label{boldH}
\end{equation}
and, in principle, this operator can have not real eigenvalues even the
ingredients $H_0$, $V_{\al}$ and three-body Hamiltonian
$H=H_{0}+\sum\limits_{\be=1}^{3}V_{\be}$ are selfadjoint operators.

In this report I will answer on the QUESTION and will give a
classification of eigenfunctions of the operator $\bH$ and its adjoint.
This report is based on refs. \cite{Ya1}, \cite{Ya2}. 
\section{Faddeev operator and its ajoint}
Let us consider the Hilbert space ${\cal H}$ of three component vectors
$F =\{ f_1, f_2, f_3\} $. The operator $\bH$ acts in ${\cal H}$ according to
the formula
\begin{equation}
(\bH F)_{\al}= H_0 f_{\al}+V_{\al}\sum_{\be}f_{\be}.
\label{fadoper}
\end{equation}
The adjoint $\bH^{*}$ is defined as 
$$
\bH^{*}=\bH_{0}+\bX\bV=
\left( 
\begin{array}{ccc}
H_0 & 0    & 0  \\
0   &H_{0} & 0  \\
0   &0     &H_0 \\
\end{array}
\right) + 
\left( 
\begin{array}{ccc}
1 & 1    & 1  \\
1 & 1    & 1  \\
1 & 1    & 1 \\
\end{array}
\right)
\left( 
\begin{array}{ccc}
V_1  & 0    & 0  \\
0    &V_2    & 0  \\
0   &0     &V_3   \\
\end{array}
\right)
$$ 
and acts as follows
\begin{equation}
(\bH^{*} G)_{\al}= H_0 g_{\al}+\sum_{\be}V_{\be}g_{\be}.
\label{adjoint}
\end{equation}
The equations for eigenvectors of operators $\bH$ and $\bH^{*}$
$$
\bH\Phi = E\Phi, \ \ \ \ \ \bH^{*}\Psi = E\Psi
$$
in components have the form
$$
H_0\vf_{\al}+V_{\al}\sum_{\be=1}^{3}\vf_{\be}=E\vf_{\al},
$$
$$
H_0\psi_{\al}+\sum_{\be=1}^{3}V_{\be}\psi_{\be}=E\psi_{\al}.
$$
The first one coincides to the Faddeev equations (\ref{fadeq}) and the
second one has the direct connection to the so called triad of
Lippmann-Schwinger equations \cite{Gloeckle}. 

It follows directly from the definitions (\ref{fadoper}) and (\ref{adjoint})
that operators $\bH$ and $\bH^{*}$ have the following invariant
subspaces:\\  
for $\bH$
$$
 {\cal H}_{s} =\{ F\in {\cal H}_{s}:\ \sum_{\al}f_{\al}=0\} ,
$$
for $\bH^{*}$
$$
{\cal H}_{p}^{*}=\{ G\in {\cal H}_{p}^{*}:\
g_{1}=g_{2}=g_{3}=g\} .
$$  
It is worth to notice that operators $\bH$ and $\bH^{*}$ on the subspaces 
${\cal H}_{s}$ and ${\cal H}_{p}^{*}$ act as free Hamiltonian $H_0$ and 
three-body Hamiltonian $H$, respectively:
$$
(\bH F)_{\al} = H_{0}f_{\al} \ \ , \mbox{if}\ \ F\in {\cal H}_{s},
$$
$$
(\bH^{*}G)_{\al}= Hg = H_{0}g+\sum_{\be}V_{\be}g \ \ , \mbox{if}\ \ G\in
{\cal H}_{p}^{*} .
$$
As a consequence the spectrum of $\bH$ on ${\cal H}_{s}$ coincides to the
spectrum of $H_{0}$ and the spectrum of $\bH^{*}$ on ${\cal H}^{*}_{}$
does to the spectrum of three-body Hamiltonian $H$. 

In order to describe eigenfunctions of operators $\bH$ and $\bH^{*}$ let
us introduce the resolvents 
$$
\bR(z)=(\bH-z)^{-1},
$$
$$
\bR^{*}(z)=(\bH^{*}-z)^{-1}.
$$
The components of these resolvents can be expressed through the resolvent
of three-body Hamiltonian and free Hamiltonian as follows
\begin{equation}
R_{\al \be}(z)=R_{0}(z)\de_{\al \be} - R_{0}(z)V_{\al}R(z),
\label{R}
\end{equation}
\begin{equation}
R^{*}_{\al \be}(z)=R_{0}(z)\de_{\al \be} - R(z)V_{\be}R_{0}(z).
\label{R*}
\end{equation}
Here
$$
R(z)=(H-z)^{-1}=(H_{0}+\sum_{\be}V_{\be}-z)^{-1},\ \ \
R_{0}(z)=(H_{0}-z)^{-1} .
$$
It is worth to note that the components of resolvents obey the following
Faddeev equations 
\begin{equation}
R_{\al \be}(z) = R_{\al}(z)\de_{\al
\be}-R_{\al}(z)V_{\al}\sum_{\ga\ne\al}R_{\ga \be}(z), 
\label{Rfad}
\end{equation}
\begin{equation}
R^{*}_{\al \be}(z) = R_{\al}(z)\de_{\al \be}-R_{\al}(z)\sum_{\ga\ne\al}
V_{\ga} R^{*}_{\ga \be}(z).
\label{R*fad}
\end{equation}
Here $R_{\al}(z)=(H_0+V_{\al}-z)^{-1}$ is the two-body resolvent for the
pair $\al$ in the three-body space.

In order to proceed it is convenient to introduce the spectral
representation for the resolvent of three-body Hamiltonian
$$
R(z)= \sum_{E_{i}}\frac{|\psi^{i}\rangle \langle \psi^{i}|}
{E_{i}-z} + 
\sum_{\ga}\int dp_{\ga}\frac{|\psi^{\ga}(p_{\ga})\rangle \langle
\psi^{\ga}(p_{\ga})|}{p^{2}_{\ga}-z} + 
\int dP \frac{|\psi^{0}(P)\rangle \langle \psi^{0}(P)|}
{P^{2}-z}.
$$
It is implied here that the system of eigenfunctions of the operator $H$
is complete {\it i.e.,}
$$
I= \sum_{i}|\psi^{i}\rangle \langle \psi^{i}|
+ 
\sum_{\ga}\int dp_{\ga}|\psi^{\ga}(p_{\ga})\rangle \langle
\psi^{\ga}(p_{\ga})|
+ 
\int dP |\psi^{0}(P)\rangle \langle \psi^{0}(P)|.
$$
Introducing this representation into (\ref{R}) and (\ref{R*}) 
one arrives to the spectral representations for components $R_{\al
\be}(z)$:
\begin{eqnarray}
R_{\al \be}(z)= \sum_{E_{i}}\frac{|\psi^{i}_{\al}\rangle \langle \psi^{i}|}
{E_{i}-z} + 
\sum_{\ga}\int dp_{\ga}\frac{|\psi^{\ga}_{\al}(p_{\ga})\rangle \langle
\psi^{\ga}(p_{\ga})|}{p^{2}_{\ga}-z} + 
\nonumber \\ 
\int dP \frac{|\psi^{10}_{\al}(P)\rangle \langle \psi^{0}(P)|}
{P^{2}-z}  
+\sum_{k=1}^{2}\int dP \frac{|u_{\al}^{k}(P)\rangle \langle w^{k}_{\be}(P)|}
{P^{2}-z}.
\label{Rsr}
\end{eqnarray}
Here $\psi^{i}_{\al}$, $\psi^{\ga}_{\al}(p_{\ga})$ and
$\psi^{10}_{\al}(P)$ are the Faddeev components of eigenfunctions of
three-body Hamiltonian:
$$
\psi^{i}_{\al}=-R_{0}(E_i)V_{\al}\psi^{i},
$$
$$
\psi^{\ga}_{\al}(p_{\ga})=
-R_{0}(\ep_{\ga}+p_{\ga}^{2}+i0)V_{\al}\psi^{\ga}(p_{\ga}),
$$
$$
\psi^{10}_{\al}(P)=\de_{\al 1} \phi^{0}(P)
-R_{0}(P^{2}+i0)V_{\al}\psi^{0}(P),
$$
where $\phi^{0}(P)$ is an eigenfunction of the free Hamiltonian:
$$
H_{0}\phi^{0}(P)=P^{2}\phi^{0}(P).
$$
A new feature in (\ref{Rsr}) is the appearance of the last term related to
the 
spurious solutions of Faddeev equations and its adjoint. The explicit
formulas for the spurious eigenfunctions $u^{k}_{\al}(P)$ of ${\bf H}$ are
of the form 
$$
u^{k}_{\al}(P)=\si^{k}_{\al}\phi^{0}(P), 
$$
where $\si_{\al}^{k}$, $k=1,2$, are the components of two noncollinear
vectors from $\bR^{3}$ lying on the plane $\sum_{\al} \si_{\al} =0$.
The spurious eigenfunctions $w^{k}_{\be}(P)$ of ${\bf H}^{*}$ can be
expressed by the formula 
$$
w^{k}_{\be}(P)= \theta^{k}_{\be}\phi^{0}(P)- 
\sum_{\alpha} [{\cal P}^{*}_{p}]_{\beta \alpha}
\theta_{\al}^{k}\phi^{0}(P),  
$$
where
$$
[{\cal P}^{*}_{p}]_{\beta \alpha}=
\sum_{i}|\psi^{i}\rangle \langle \psi^{i}_{\al}|
+ 
\sum_{\ga}\int dp^{'}_{\ga}|\psi^{\ga}(p^{'}_{\ga})\rangle \langle
\psi^{\ga}_{\al}(p^{'}_{\ga})|
+ 
\int dP^{'} |\psi^{0}(P^{'})\rangle \langle \psi^{01}_{\al}(P^{'})|.
$$ 
Here the vectors $\theta^{k}\in \bR^{3}$ are defined by following
biorthogonality conditions
$$
\sum_{\alpha}\theta_{\al}^{i}\sigma_{\al}^{j}=\de_{ij},\ \ \  i,j=0,1,2,
$$
with $\si^{0}_{\al}=\de_{\al 1}$ and $\theta_{\al}^{0}=1$.

For the components of resolvent $R^{*}_{\al \be}(z)$ one can obtain the
similar to (\ref{Rsr}) formula 
$$
R^{*}_{\al \be}(z)= \sum_{E_{i}}\frac{|\psi^{i}\rangle \langle
\psi^{i}_{\be}|} 
{E_{i}-z} + 
\sum_{\ga}\int dp_{\ga}\frac{|\psi^{\ga}(p_{\ga})\rangle \langle
\psi^{\ga}_{\be}(p_{\ga})|}{p^{2}_{\ga}-z} + 
\int dP \frac{|\psi^{0}(P)\rangle \langle \psi^{10}_{\be}(P)|}
{P^{2}-z}  +
$$
\begin{equation}
+\sum_{k=1}^{2}\int dP \frac{|w_{\al}^{k}(P)\rangle \langle u^{k}_{\be}(P)|}
{P^{2}-z}.
\label{R*sr}
\end{equation}

It is follows from (\ref{Rsr}) and (\ref{R*sr}) that operators $\bH$ and
$\bH^{*}$ have the following system of eigenfunctions:\\ 
 
$\{ $ $\Phi^{i}$, $\Phi^{\ga}(p_{\ga})$, $\Phi^{10}(P)$ and $U^{k}(P)$ $\} $
$$
\bH\Phi^{i}=E_{i}\Phi^{i},
$$
$$
\bH\Phi^{\ga}(p_{\ga})=(\ep_{\ga}+p_{\ga}^{2})\Phi^{\ga}(p_{\ga}),
$$
$$
\bH\Phi^{10}(P)=P^{2}\Phi^{10}(P),
$$
$$
\bH U^{k}(P)=P^{2}U^{k} , \ \ k=1,2;
$$

$\{$ $\Psi^{i}$, $\Psi^{\ga}(p_{\ga})$, $\Psi^{10}(P)$ and $W^{k}(P)$ $\} $
$$
\bH^{*}\Psi^{i}=E_{i}\Psi^{i},
$$
$$
\bH^{*}\Psi^{\ga}(p_{\ga})=(\ep_{\ga}+p_{\ga}^{2})\Psi^{\ga}(p_{\ga}),
$$
$$
\bH^{*}\Psi^{10}(P)=P^{2}\Psi^{10}(P),
$$
$$
\bH^{*} W^{k}(P)=P^{2}W^{k} , \ \ k=1,2, 
$$
with components of physical eigenfunctions: 
$$
\phi^{i}_{\al}=-R_{0}(E_i)V_{\al}\psi^{i},
$$
$$
\phi^{\ga}_{\al}(p_{\ga})=
-R_{0}(\ep_{\ga}+p_{\ga}^{2}+i0)V_{\al}\psi^{\ga}(p_{\ga}),
$$
$$
\phi^{10}_{\al}(P)=\de_{\al 1} \phi^{0}(P)
-R_{0}(P^{2}+i0)V_{\al}\psi^{0}(P),
$$
for $\bH$ and with components for physical eigenfunctions: 
$$
\psi^{i}_{\al}=\psi^{i},
$$
$$
\psi^{\ga}_{\al}(p_{\ga})=
\psi^{\ga}(p_{\ga}),
$$
$$
\psi^{10}_{\al}(P)=\psi^{0}(P)
$$
for $\bH^{*}$.

Physical eigenfunctions span the physical subspace of ${\cal H}$. This
subspace can be defined as 
$$
{\cal H}_{p} = {\cal P}_{p}{\cal H},
$$
where the projection ${\cal P}_{p}$ is defined by formula
$$
{\cal P}_{p}=
\sum_{i}|\Phi^{i}\rangle \langle \Psi^{i}|
+ 
\sum_{\ga}\int dp_{\ga}|\Phi^{\ga}(p_{\ga})\rangle \langle
\Psi^{\ga}(p_{\ga})|
+ 
\int dP |\Phi^{10}(P)\rangle \langle \Psi^{10}(P)|.
$$
Spurious solutions span the spurious subspace of ${\cal H}$:
$$
{\cal H}_{s}= {\cal P}_{s}{\cal H}.
$$
where 
$$
{\cal P}_{s} = \sum_{k=1}^{2} \int dP |U^{k}(P)\rangle \langle W^{k}(P)|.
$$
It is follows from construction and completeness of eigenfunctions of
three-body Hamiltonian, that physical and spurious subspaces are complete
in ${\cal H}$:
$$
{\cal H}= {\cal H}_{p}+{\cal H}_{s}.
$$
The same is valid for physical and spurious subspaces of operator
$\bH^{*}$: 
 
$$
{\cal H}= {\cal H}_{p}^{*}+{\cal H}_{s}^{*},
$$
where the subspaces ${\cal H}_{p}^{*}$ and ${\cal H}_{s}^{*}$ are defined
as 
$$
{\cal H}_{p}^{*}= {\cal P}_{p}^{*}{\cal H}, \ \ \ 
{\cal H}_{s}^{*}= {\cal P}_{s}^{*}{\cal H}.
$$
Here the operators ${\cal P}_{p}^{*}$ and ${\cal P}_{s}^{*}$ are Hilbert
space adjoints for ${\cal P}_{p}$ and ${\cal P}_{s}$.

\noindent 
The results described above can be summarized as the following \\
{\bf Theorem}:  {\it Faddeev operator} {\bf H} 
$$
\bH= 
\left( 
\begin{array}{ccc}
H_0 & 0    & 0  \\
0   &H_{0} & 0  \\
0   &0     &H_0 \\
\end{array}
\right) + 
\left( 
\begin{array}{ccc}
V_1  & 0    & 0  \\
0    &V_2    & 0  \\
0   &0     &V_3   \\
\end{array}
\right)
\left( 
\begin{array}{ccc}
1 & 1    & 1  \\
1 & 1    & 1  \\
1 & 1    & 1 \\
\end{array}
\right)
$$
{\it and its adjoint} ${\bf H}^{*}$ 
$$
\bH^{*}=
\left( 
\begin{array}{ccc}
H_0 & 0    & 0  \\
0   &H_{0} & 0  \\
0   &0     &H_0 \\
\end{array}
\right) + 
\left( 
\begin{array}{ccc}
1 & 1    & 1  \\
1 & 1    & 1  \\
1 & 1    & 1 \\
\end{array}
\right)
\left( 
\begin{array}{ccc}
V_1  & 0    & 0  \\
0    &V_2    & 0  \\
0   &0     &V_3   \\
\end{array}
\right)
$$ 
{\it have coinciding spectrums of real eigenvalues}
$$
\sigma({\bf H})=\sigma({\bf H}^{*})=\sigma(H)\cup \sigma(H_{0}), 
$$
{\it where the physical part of the spectrum} $\sigma(H)$ {\it is the
spectrum of the three-body 
Hamiltonian} $H=H_{0}+\sum_{\alpha}V_{\alpha}$ {\it and the spurious part}
$\sigma(H_{0})$ 
{\it is the spectrum of the free Hamiltonian} $H_{0}$.   
{\it The sets of physical and spurious eigenfunctions are complete and
biorthogonal in the sense:}
$$
{\cal P}_{p}+{\cal P}_{s}={\cal P}^{*}_{p}+{\cal P}^{*}_{s}=I,
$$
$$
{\cal P}^{2}_{p(s)}={\cal P}_{p(s)}, \ \ {{\cal P}^{*}}^{2}_{p(s)}= {{\cal
P}^{*}}_{p(s)}, \ \   
{\cal P}_{p}{\cal P}^{*}_{s}=0, \ \ {\cal P}_{s}{\cal P}^{*}_{p}=0.
$$
\section{Extension on CCA equations}
It is shown that the matrix operator generated by Faddeev equations in
differential form has the (additional to physical) spurious spectrum. The
existence of this spectrum strongly relates to the invariant spurious
subspace formed by components which sum is equal to zero. The theorem
formulated in preceding section can be extended on any matrix operator
corresponding to few-body equations for components of wave-function
obtained in framework of so called coupled channel array (CCA) method
\cite{Levin} as follows.  CCA equations can be written in the matrix form
as 
\begin{equation}
{\bf H}\Phi = E \Phi,  
\label{CCA}
\end{equation}
where ${\bf H}$ is a $n\times n$ matrix operator acting in the Hilbert
space ${\cal H}$ of vector-functions $\Phi $ with components $\phi_{1},
\phi_{2},..., \phi_{n}$ each belonging to few-body system Hilbert space $h$. 
The equivalence of Eq. (\ref{CCA}) to the Schr\"{o}dinger equation 
$H\psi=(H_{0}+\sum_{\beta}V_{\beta})\psi=E\psi$ by requiring
$\sum_{\alpha}\phi_{\alpha}=\psi$ can be reformulated as the following
intertwining property for operators ${\bf H}$ and $H$
\begin{equation}
{\cal S}{\bf H} = H {\cal S}.
\label{SHHS}
\end{equation}
Here ${\cal S}$ is the summation operator 
$$
{\cal S}\Phi = \sum_{\alpha}\phi_{\alpha}
$$
acting from ${\cal H}$ to $h$. Due to (\ref{SHHS}) the subspace ${\cal
H}_{s}$ formed by spurious vectors such that ${\cal S}\Phi =0$ is
invariant with respect to ${\bf H}$ and as a consequence the operator
${\bf H}$ has the spurious spectrum $\sigma_{s}$. Clearly, that the
concrete form of $\sigma_{s}$ and of corresponding eigenfunctions depends
on the particular form of the matrix operator ${\bf H}$ and is the
subject of special investigation.     

The physical part $\sigma_{p}$ of the spectrum of ${\bf H}$ can be found
with adjoin variant of (\ref{SHHS})
\begin{equation}
{\bf H}^{*}{\cal S}^{*}= {\cal S}^{*}H,
\label{SHHS*}
\end{equation}
where adjoint ${\cal S}^{*}$ acts from $h$ to ${\cal H}$ according to the
formula 
$$
[{\cal S}^{*}\phi]_{\alpha}= \phi.
$$ 
It follows from Eq. (\ref{SHHS*}) that the range ${\cal H}^{*}_{p}$ of
operator ${\cal S}^{*}$ consisting of vector-functions with the same
components 
is invariant with respect to ${\bf H}^{*}$ and the restriction of ${\bf
H}^{*}$ on ${\cal H}^{*}_{p}$ is reduced to few-body Hamiltonian $H$. So
that, $\sigma_{p}=\sigma(H)$ and, similarly to the case of the Faddeev
operator, the same formula 
for the spectrums of operators ${\bf H}$ and ${\bf H}^{*}$ is valid 
$$
\sigma({\bf H})=\sigma({\bf H}^{*})= \sigma(H)\cup \sigma_{s},
$$
where $\sigma(H)$ is the spectrum of few-body Hamiltonian
$H=H_{0}+\sum_{\alpha}V_{\alpha}$.    
 
\section*{Acknowledgement}
This work was partially supported by Russian Foundation for Basic Research
grant No. 98-02-18190. Author is grateful to Organizing Committee of 16
European Conference on Few-Body Problem in Physics for financial support
of his participation in the Conference. 


\begin{thebibliography}{}
\bibitem{NF} H.P. Noyes, H. Fiedelday: In: {\it Three-particle scattering
in quantum mechanics}, p. 195.  New-York-Amsterdam 1968 
\bibitem{FM} L.D. Faddeev, S.P. Merkuriev: {\it Quantum Scattering Theory
for several particle systems}. Doderecht: Kluwer Academic Publishers 1993
\bibitem{Friar} J.L. Friar, B.F. Gibson, G.L. Paine: Phys. Rev. 
{\bf C22}, 284 (1980)
\bibitem{Pup1} V.V. Pupyshev: Theor. Math. Phys. {\bf 81}, 86 (1989) 
\bibitem{Pup2} V.V. Pupyshev: Phys. Lett. {\bf A140}, 284 (1989)
\bibitem{RYa} V.A. Roudnev, S.L. Yakovlev: Phys. of Atom. Nuclei {\bf 58}, 
 1662 (1995) 
\bibitem{Ya1} S.L. Yakovlev: Theor. Math. Phys. {\bf 102}, 
323 (1995)
\bibitem{Ya2} S.L. Yakovlev: Theor. Math. Phys. {\bf 107}, 
513 (1996)
\bibitem{Gloeckle} W. Gl\"{o}ckle: Nucl. Phys. {\bf A141}, 620 (1970) 
\bibitem{Levin} F.S. Levin: Ann. Phys. (N.Y.) {\bf 130}, 139 (1980)
\end{thebibliography}
\end{document}